\documentclass[aps,prl,letterpaper,10pt,twocolumn,showpacs,superscriptaddress]{revtex4-1}

\usepackage[utf8]{inputenc}
\usepackage{amsfonts,amssymb,amsmath,amsthm,graphicx}
\usepackage{url}
\usepackage{dsfont}
\usepackage{bbm}
\usepackage[usenames,dvipsnames]{xcolor}
\usepackage{nicefrac}

\usepackage[T1]{fontenc}
\usepackage[english]{babel}
\makeatletter
\let\it@comma@def\active@comma
\makeatother

\usepackage{natbib}
\usepackage{setspace}
\usepackage{subfigure}
\usepackage{comment}
\usepackage[normalem]{ulem} 
\usepackage{todonotes}
\usepackage[colorlinks = true, urlcolor  = blue]{hyperref}

\def\ket #1{\vert #1\rangle}

\newcommand{\beq}{\begin{equation}}
\newcommand{\eeq}{\end{equation}}

\newcommand{\degrees}{^{\circ}}
\newcommand*\rfrac[2]{{}^{#1}\!/_{#2}}

\begin{document}

\title{Extending Wheeler's delayed-choice experiment to Space
}
\author{Francesco Vedovato}
\affiliation{Dipartimento di Ingegneria dell'Informazione, Universit\`a degli Studi di Padova, Padova, Italy}
\author{Costantino Agnesi}
\affiliation{Dipartimento di Ingegneria dell'Informazione, Universit\`a degli Studi di Padova, Padova, Italy}
\author{Matteo Schiavon}
\affiliation{Dipartimento di Ingegneria dell'Informazione, Universit\`a degli Studi di Padova, Padova, Italy}
\author{Daniele Dequal}
\affiliation{Dipartimento di Ingegneria dell'Informazione, Universit\`a degli Studi di Padova, Padova, Italy}
\affiliation{Matera Laser Ranging Observatory, Agenzia Spaziale Italiana, Matera, Italy}
\author{Luca Calderaro}
\affiliation{Dipartimento di Ingegneria dell'Informazione, Universit\`a degli Studi di Padova, Padova, Italy}
\author{Marco Tomasin}
\affiliation{Dipartimento di Ingegneria dell'Informazione, Universit\`a degli Studi di Padova, Padova, Italy}
\author{Davide Giacomo Marangon}
\affiliation{Dipartimento di Ingegneria dell'Informazione, Universit\`a degli Studi di Padova, Padova, Italy}
\author{Andrea Stanco}
\affiliation{Dipartimento di Ingegneria dell'Informazione, Universit\`a degli Studi di Padova, Padova, Italy}
\author{Vincenza Luceri}
\affiliation{e-GEOS spa, Matera, Italy}
\author{Giuseppe Bianco}
\affiliation{Matera Laser Ranging Observatory, Agenzia Spaziale Italiana, Matera, Italy}
\author{Giuseppe Vallone}
\affiliation{Dipartimento di Ingegneria dell'Informazione, Universit\`a degli Studi di Padova, Padova, Italy}
\author{Paolo Villoresi}
\email{paolo.villoresi@dei.unipd.it}
\affiliation{Dipartimento di Ingegneria dell'Informazione, Universit\`a degli Studi di Padova, Padova, Italy}
\date{\today}



\begin{abstract}
Gedankenexperiments have consistently played a major role in the development of quantum theory.
 A paradigmatic example is  Wheeler's delayed-choice experiment, a wave-particle duality test 
 that cannot be fully understood using only classical concepts.
Here, we implement  Wheeler's idea along a satellite-ground interferometer which extends for  thousands of kilometers in Space. We exploit temporal and polarization degrees of freedom of photons reflected by a fast moving satellite equipped with retro-reflecting mirrors. We observed the complementary wave-like or particle-like behaviors at the ground station by choosing the measurement apparatus while the photons are propagating from the satellite to the ground.
Our results confirm quantum mechanical predictions, demonstrating the need of the dual wave-particle interpretation, at this unprecedented scale. Our work paves the way for novel applications of quantum mechanics in Space links involving multiple photon degrees of freedom.
\end{abstract}

\maketitle

{\it Introduction.\textemdash}
Quantum communications in Space 
enable the investigation of the basic 
principles of quantum mechanics in a radically new scenario. As envisioned in theoretical works \cite{Aspelmeyer2003,Tomaello2011,ride12cqg,Bourgoin2013,Bruschi2014}
and satellite mission proposals~\cite{Scheidl2013,Jennewein2014, Pan2014, SpaceQuest}, 
quantum information protocols \cite{NielsenChuang,Gisin2007}
have
breached the Space frontier~\cite{NatureEditorial} 
 in recent  experimental demonstrations  \cite{Villoresi2008,Yin2013,vall15prl,dequ15prl,Vallone_prl_2016,Tang_2016,Gunther2016}. 
These developments foster the implementation in Space of
fundamental tests of Physics 
such as the \emph{Gedankenexperiments} which 
highlight the counterintuitive aspects of quantum theory.

These thought experiments played a primary role in the famous debate between Einstein and Bohr \cite{Bohr1949}, concerning the completeness of quantum mechanics~\cite{Einstein1935,Bohr1935} and the concept of \emph{complementarity}~\cite{Bohr1928}. 
The most disturbing implication of complementarity is the \emph{wave-particle duality} of quantum matter, that is  the impossibility of revealing at the same time both the wave-like and particle-like properties of a quantum object. Bohr pointed out that it is necessary to consider the whole apparatus in order to determine which property is measured, stating that there is no difference ``whether our plans of constructing or handling the instruments are fixed beforehand or whether we postpone the completion of our planning until a later moment''~\cite{Bohr1949}.

\begin{figure}[t]
\centering
\includegraphics[width=8cm]{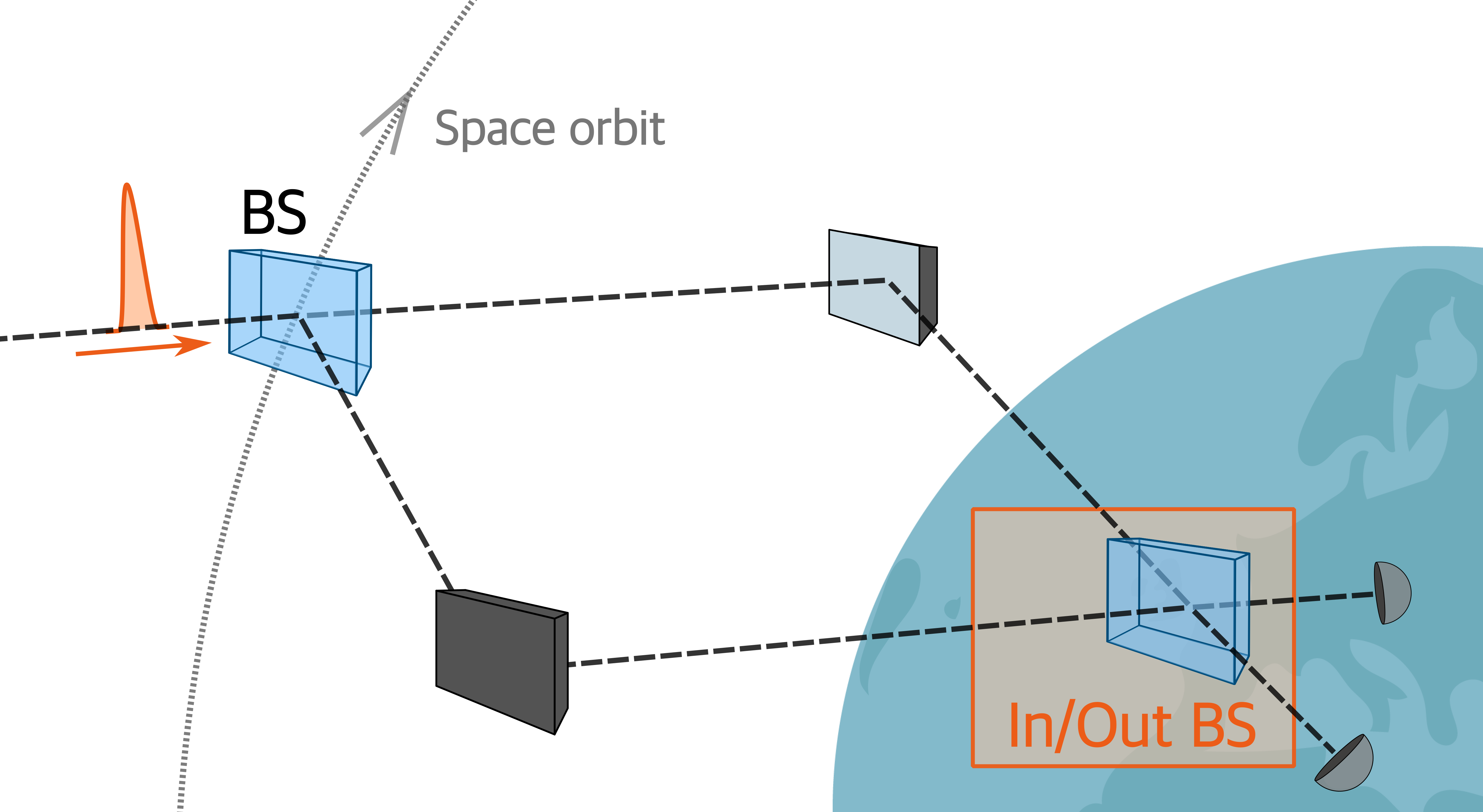}
\caption{{\it Pictorial representation of Wheeler's delayed choice experiment in Space.} 
A photon wave packet enters the first beam splitter (BS) of an interferometer which extends along thousands of kilometers in Space.  
The interferometer can be randomly arranged according to two configurations that correspond 
to the presence or absence of the second beam splitter (In/Out BS) located on  Earth.
Following Wheeler's idea, the configuration choice is performed when the photon has already entered the interferometer. }
\label{fig:1}
\end{figure}

John Wheeler pushed this observation to the extreme and conceived his \emph{delayed-choice Gedankenexperiment} to highlight the contradictory interpretation given by classical physics~\cite{Wheeler1978, Wheeler1984}. 
{In his idea, a photon emerging from the first beam-splitter (BS) of a Mach-Zehnder interferometer (see Figure~\ref{fig:1}) may find two alternative configurations.
Given the presence or absence of a second BS at the output of the interferometer, the apparatus measures the wave-like or particle-like character of the photon.
Indeed, if the BS is absent  only one of the two detectors will fire, reflecting the fact that the photon traveled along only one arm of the interferometer
and revealing \emph{which-path}  it took,
as a classical particle would have done.
If the BS is present, 
\emph{interference}
can be observed,
reflecting the fact that the photon traveled both routes, as a 
classical wave would have done. 

If the configuration is chosen after the entrance of the photon into the interferometer, 
a purely classical interpretation of the process in which the photon decides its nature at the first BS would imply a seeming violation of causality.
On the other hand, in the quantum mechanical interpretation of the experiment, 
the photon maintains its dual wave-particle nature until the very end of the experiment, when it is detected.

Here, we extend the delayed-choice paradigm to Space, 
as sketched in  Figure~\ref{fig:1}, 
by combining temporal and polarization degrees of freedom of photons reflected by a rapidly moving  satellite in orbit.
In our scheme, the two 
 paths of the interferometer
are represented by two time-bins, each with orthogonal polarization.
We are able to demonstrate the need of  the  dual wave-particle model for a propagation distance up to 3500 km,
demonstrating the validity of the quantum mechanical description at a much larger 
scale than all previous experiments.

So far, several implementations of Wheeler's \emph{Gedankenexperiment} have been realized on ground (see~\cite{Jacques2007} for the realization closest to the original idea and \cite{Ma_review_2016} for a complete review).
An alternative way of interpreting the delayed choice experiment is within the framework of {\it quantum-erasure}  \cite{Scully_1982,Scully_1991}.
Furthermore, a quantum delayed choice version of the experiment, where a quantum ancilla controls the second BS, has been recently proposed \cite{Ionicioiu_2011} and  realized \cite{Tang_2012,Peruzzo_2012}.

\bigskip
{\it Description of the experiment.\textemdash}
We realized the 
experiment at the Matera Laser Ranging Observatory (MLRO)
 of the Italian Space Agency.
 At MLRO, we have already tested the feasibility of receiving qubits encoded in the polarization of single photons
 \cite{vall15prl} and of observing interference between two temporal modes throughout satellite-ground channels
in \cite{Vallone_prl_2016}.
A pulsed laser (100 MHz repetition rate, $\lambda = 532$ nm wavelength, 
$\sim1$ nJ energy per pulse), diagonally polarized 
and paced by an atomic clock, enters into an unbalanced Mach-Zehnder interferometer (MZI), as sketched in Figure \ref{fig:2}. 
The combined action of the first polarizing beam splitter (MZI-PBS) and of the unbalance of the MZI transforms each laser pulse into a 
superposition of two temporal and polarization modes.
In fact, the long arm of the MZI is traveled by the vertically polarized component of the beam while the horizontally polarized component travels along the short arm.  
The separation between the two temporal modes is about $\Delta t \approx 3.5$ ns (see the Appendix for more details).

\begin{figure}[t!]
\centering
\includegraphics[width=8cm]{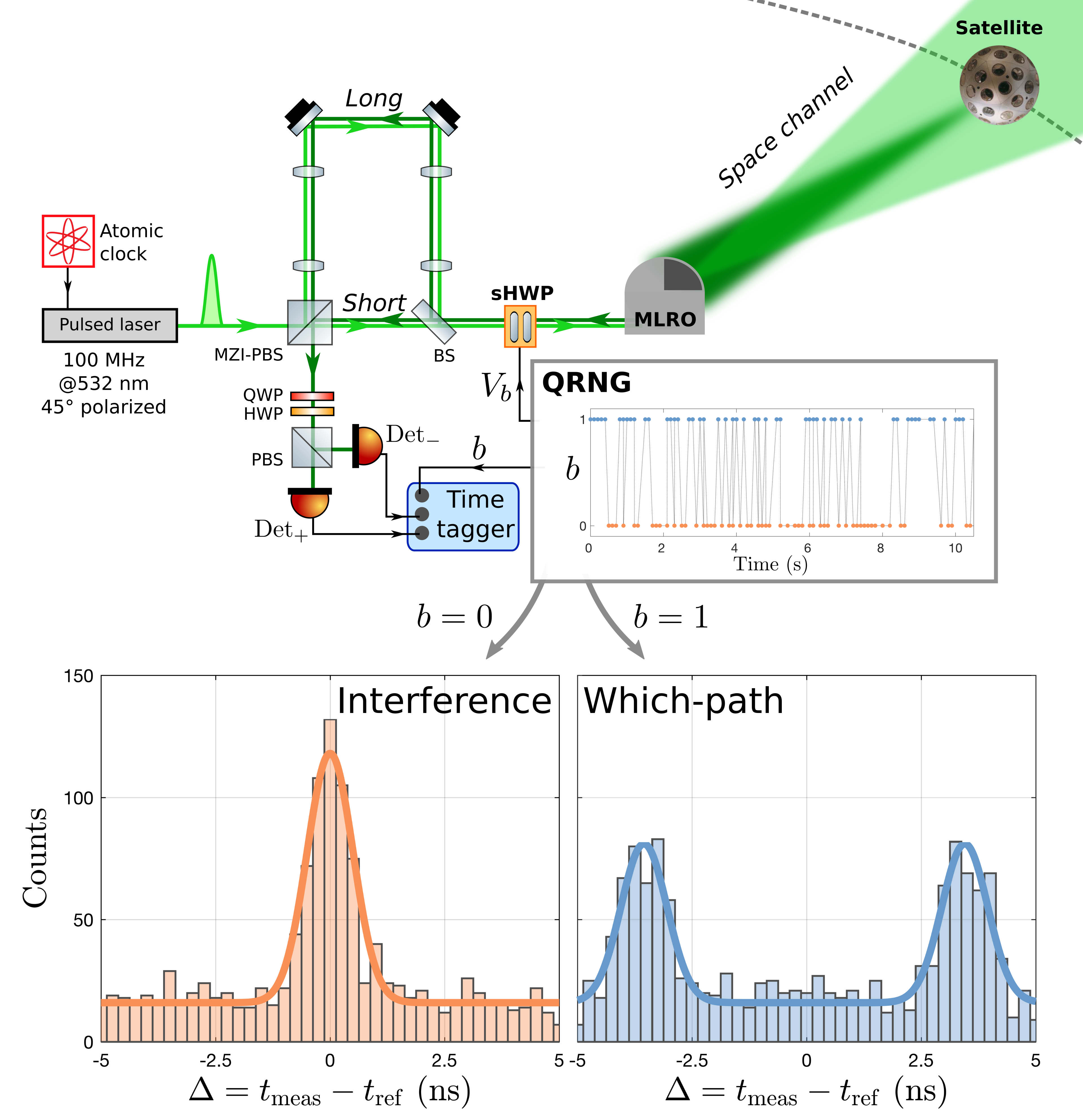}
\caption{{\it Scheme of the experimental setup and detection histograms.} A pulsed laser synchronized with the MLRO atomic clock exits the MZI in two temporal and polarization modes. The sHWP
leaves the pulses unperturbed and the telescope directs the beam to a target satellite. After the reflection, the photons are collected at the ground by the same telescope and injected into the optical table. The photons pass through the sHWP whose behavior is set according to the bit $b$ extracted from a QRNG. 
In the inset, a sample of the extracted bits relative to 10 seconds of detections is shown.
At the MZI output, two waveplates, a PBS and two single photon detectors 
perform a polarization measurement in the $\lbrace \ket{+}, \ket{-}\rbrace$ basis. 
According to the value $b$ of the random bit, interference or which-path measurement is performed, as shown by the  detection histograms
for a passage of the Starlette satellite. The counts in the central peak on the left histogram are comparable to the sum of the counts associated to the lateral peaks on the right one, as expected.}
\label{fig:2}
\end{figure}

The pulses then pass through 
two liquid crystal retarders (LCRs) whose combined action is equivalent to a single switchable ({\bf on/off}) half waveplate (sHWP) inclined at $45\degrees$ with respect to the fast axis. During the transmission period, the sHWP is always {\bf off}, leaving the outgoing beam unperturbed.
The light is then directed to a target satellite equipped with polarization maintaining corner-cube retroreflectors via a telescope \cite{vall15prl}.
The corner cubes of the target satellite redirect the beam back to the ground station.
Furthermore, the radial motion of the satellite introduces a kinematic phase shift between the two time-bins given by $\varphi(t) = \frac{2 \beta(t)}{1 + \beta(t)} \frac{2\pi c \Delta t}{\lambda}$ where $\beta(t) = v_r(t)/c$ with $v_r(t)$ the instantaneous satellite radial velocity with respect to the ground and $c$ the speed of light in vacuum, as previously demonstrated by our group in \cite{Vallone_prl_2016}.
The photons returning from the satellite are collected by the same telescope and injected into the optical table where they 
re-encounter the same sHWP and the MZI. At an exit port of the MZI-PBS (see Fig.~\ref{fig:2}),  we 
perform a polarization measurement in the diagonal and anti-diagonal basis 
$\lbrace \ket{+}, \ket{-} \rbrace$ with $\ket{\pm} = (\ket{H} \pm \ket{V})/\sqrt{2}$.

While the photons are propagating back to the ground station, a quantum random number generator (QRNG) extracts a random bit $b \in \lbrace 0, 1 \rbrace$ with 50:50 probability. The QRNG is based on the time of arrival of single photons in attenuated light and 
its performances will be detailed in \cite{randy}.
The bit value sets the  voltages $V_b$ applied to the LCRs, determining the {\bf on} or {\bf off} behavior of the sHWP. 
The latter determines whether we perform a measurement that reveals the particle-like (sHWP  {\bf on}) or wavelike-like (sHWP {\bf off}) 
behavior of the photons returning from the satellite. 
The random bits are generated while the photons are traveling from the satellite to the ground station, ensuring space-like separation between the measurement choice and the last interaction with the  apparatus, i.e. the reflection by the satellite  (as it will be detailed in the following).

Let us first suppose that the QRNG extracts a $b = 0$ bit causing the sHWP to remain \textbf{off}, 
leaving the photon unchanged as it re-enters the MZI.
At the exit port of the MZI-PBS towards the detectors in Fig.~\ref{fig:2}, 
only the horizontally polarized component that propagated through the long arm
 and the vertically polarized component that traveled along the short arm can be detected.
Since this is the reverse situation compared to the outward passage through the MZI, 
the two polarization modes will recombine into a single temporal mode, 
loosing all which-path information and allowing us to observe a $\varphi$-dependent interference, which is the fingerprint of the wave-like nature of the photon.
Indeed, in this case the probabilities of a click in the detectors $\mathrm{Det}_{\pm}$  are
given by
\begin{equation}
P_{\pm}^{b=0}(t) = \frac{1}{2} \left[ 1 \pm \mathcal{V} (t)\cos{\varphi(t)} \right] \label{eq_interference}
\end{equation}
where $\mathcal{V}(t) \approx 1$ is the theoretical visibility as in \cite{Vallone_prl_2016}.

Let us now suppose that the QRNG extracts a $b = 1$ bit, switching  the sHWP \textbf{on} and
swapping the horizontal and vertical polarizations before the photon re-enters the MZI.
The polarization transformation causes each component of the state to re-travel along the same arm compared to the outwards passage through the MZI. As a result, the photon can be detected at two distinct times  separated by 2$\Delta t$ (with  50\% probability for each detector $\mathrm{Det}_{\pm}$,  i.e. $P_{\pm}^{b=1}(t) = 1/2$), giving which-path information and evidencing the particle nature of the photon. 

\bigskip
{\it Implementation of the delayed-choice.\textemdash}  
Simultaneous tracking of the target satellite via standard laser ranging technique (SLR) allows the determination with few tens of picosecond accuracy of the photon's time of flight, or \emph{round trip time} ($\mathrm{rtt}$). Furthermore, SLR allows an accurate estimation of the satellite radial velocity,
which is crucial for the determination of the kinematic phase $\varphi(t)$.
The laser ranging technique exploits a bright laser signal with pulses at a 10 Hz repetition rate, 
synchronized with the 100 MHz train used in the experiment (see the Appendix for more details).  

\begin{figure}[t]
\centering
\includegraphics[width=8cm]{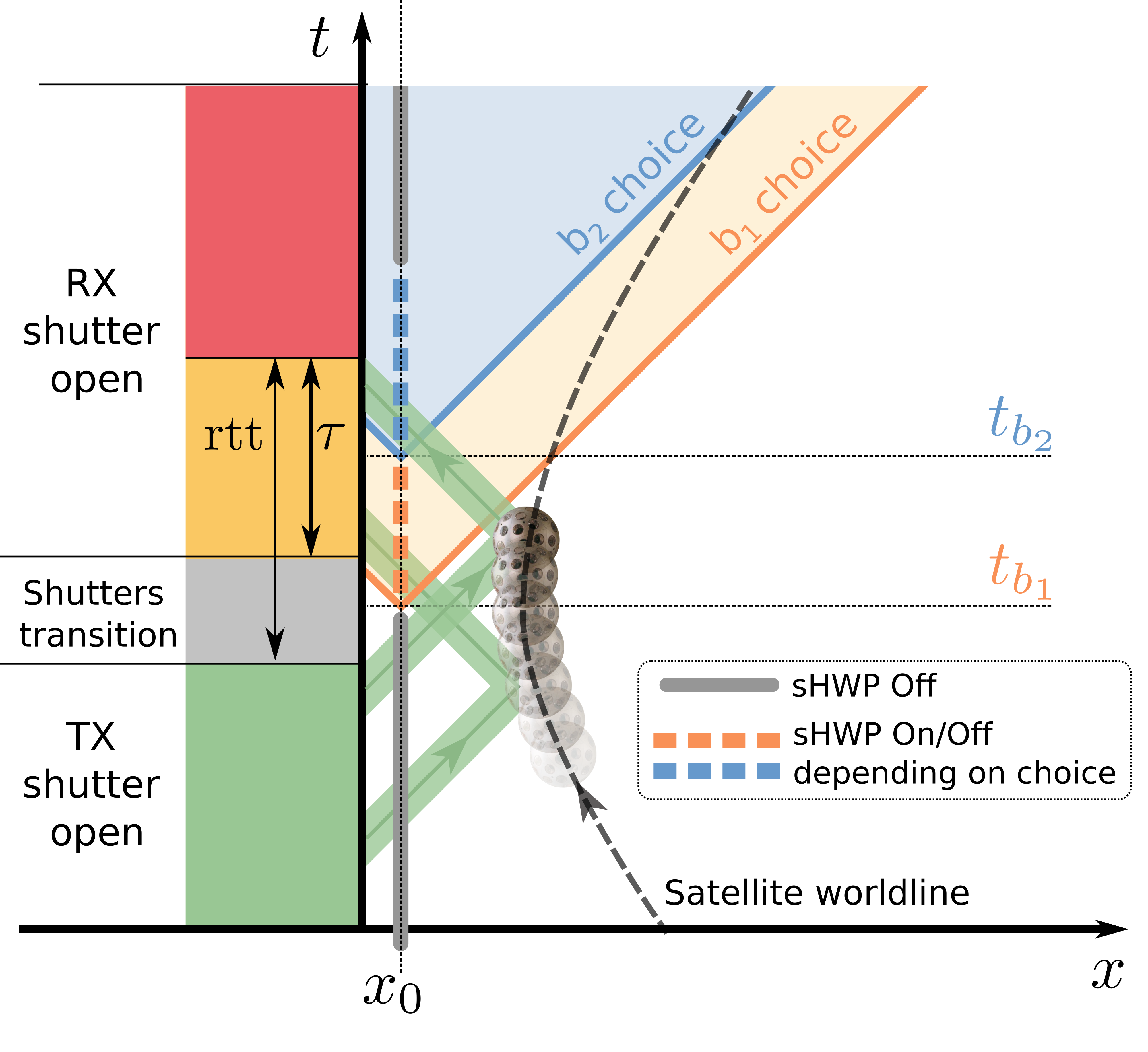}
\caption{\label{fig:3}{\it Minkowski diagram of the experiment.} 
Along the temporal axis (not to scale) a 100 ms cycle between two SLR pulses is represented.
The $x$-axis represents the radial coordinate (not to scale) from the detectors,
where $x_0$ is the position of {both} the switchable HWP and the QRNG.
The dotted line is the satellites worldline.
As detailed in the main text, we only considered the detections in the temporal window $\tau$.
A fast FPGA controller synchronized in real time with the MLRO tracking system drives the two shutters and the QRNG. 
For each cycle, we perform two independent measurements via the random bit extracted by the QRNG at times $t_{b_1}$  and $t_{b_2}$,
causally disconnected from the photon reflection at the satellite.
The cycle is repeated for each 100 MHz train between two SLR pulses. }
\end{figure}

We separated each 100 ms cycle between two subsequent SLR pulses in two periods by using two mechanical shutters (see the Figure \ref{fig:3}). In the first half of the 100 ms, only the transmitting shutter (TX shutter) is open, while the receiving one (RX shutter) is closed. In the second half of the time slot, 
 the TX shutter is closed while the RX shutter is open and the detectors can receive the photons coming from the satellite. 
Furthermore, since the shutters require a certain time to open and close completely, the effective detection time period is limited by 
the shutters transition time ($t_{\mathrm{trans}} \sim 5$ ms), as sketched in the figure. So, there exist a precise temporal window 
$\tau = \mathrm{rtt} - t_{\mathrm{trans}}$ where we expect to receive photons from the satellite. 
 The value of $\tau$ depends on the actual $\mathrm{rtt}$ which is continuously changing along the satellite orbit. However, the SLR technique described above allows the transmission and reception phases of the protocol to be synchronized in real time by using a fast Field Programmable Gate Array (FPGA) controller.

A faithful realization of Wheeler's  experiment requires that the entrance of the photon in the interferometer is not in the future light-cone of the measurement choice.
Moreover, the latter must be realized in a random manner: 
 this prevents any causal influence of the measurement 
 choice on the behavior of the photon. 
 
Our implementation is performed over a Space channel with length of the order of thousands of kilometers, 
corresponding to a $\mathrm{rtt}$ of the order of 10 ms. 
We designed the experiment to guarantee that the choice of the measurement apparatus 
is space-like separated from the reflection of the photon from the satellite, as shown in the Minkowski diagram of Figure~\ref{fig:3}.
This guarantees that, in a purely classical interpretation, a photon 
``should have decided its nature'' at most  just after its reflection from the satellite.

For each cycle, we performed two independent choices that will affect the detections in the acceptable temporal window $\tau$ by driving the QRNG with the same FPGA controller used for the shutters. The sHWP behavior at the photon return is set according to the bits $b_1$ and $b_2$ extracted by the QRNG. 
The first choice is performed at $t_{b_1}$, corresponding to the middle of the shutters transition phase. 
The second choice is at $t_{b_2}$, which occurs with a delay 
$\mathrm{rtt}/2$ with respect to the first choice. 
The detected photons are divided into two groups, each characterized by a value of the bit choice.
In this way, all the photons of a given group were already reflected by the satellite when the corresponding bit choice was performed.

\bigskip
{\it Experimental results.\textemdash}
We selected the passages of two low-Earth-orbit (LEO) satellites equipped with  polarization maintaining corner-cube retroreflectors, namely Beacon-C dated November 1st, 2016 h 23.18 CEST (with slant distance ranging from 1264 to 1376 km with respect to the MLRO Observatory) and Starlette dated
November 1st, 2016 h 22.00 CEST (slant distance  ranging from 1454 to 1771 km). 

The synchronization between our signal and the bright laser ranging pulses allowed us to predict  the expected time of arrival $t_{\mathrm{ref}}$ of the photons, which is not periodic along the orbit due to the satellite motion. The effective time of arrival $t_{\mathrm{meas}}$ were tagged by a time-to-digital converter (time-tagger of Fig.~\ref{fig:2}).  
Therefore, we may obtain a detection histogram as a function of the time difference $\Delta = t_{\mathrm{meas}} - t_{\mathrm{ref}}$, 
as shown in the bottom panel of Figure~\ref{fig:2} for the data recorded in the $\mathrm{Det}_-$ detector in the passage of Starlette satellite 
(results for the $\mathrm{Det}_+$ are analogous). 

As previously described, we separated the detections in two groups according to the setting of the sHWP. 
In Fig.~\ref{fig:2}, on the left histogram we gathered all the detections characterized by the bit value $b = 0$ 
and we obtain a single central peak where which-path information is erased and the interference effects should be observed.
The peak width is determined mostly by the timing jitter of the detector which is about $0.5$ ns RMS.
On the right histogram, the extracted bit $b$ was equal to 1, 
and we obtain a histogram with two well separated lateral peaks, manifesting the expected particle-like behavior. An indication of good assessment for the setup is given by the fact that the peak obtained when $b=0$  is comparable with the sum of the two lateral peaks obtained when $b=1$, as the number of ``0'' and ``1'' bits from the QRNG is balanced. We note that, even if interference is expected in the $b=0$ case, it is not apparent in Fig.~\ref{fig:2}, since we are not taking into account the phase shift $\varphi(t)$ introduced by the satellite, and thus the interference effect is completely averaged over all the data. 

To evaluate the role of the kinematic phase $\varphi(t)$, these two data sets were further separated into ten phase intervals of length $\pi/5$ rads and defined by {$\mathcal{I}_j \equiv \left[ (2j-1)\rfrac{\pi}{10}, (2j+1)\rfrac{\pi}{10} \right]$ with $j = 0,\dots,9$.} For each phase interval we selected the detection events characterized by $\varphi \ (\mathrm{mod} \ 2\pi) \in \mathcal{I}_j$.
Then, for each selected data set  we evaluated the  detection histogram as a function of the time difference $\Delta$, as described above. By using these histograms we determined the relative frequency of detection 
$f_{\pm} = \frac{N_{\pm}}{N_+ + N_-}$,
where $N_{\pm}$ are the counts associated to the detections recorded by $\mathrm{Det}_{\pm}$. $N_{\pm}$ is estimated by taking the integral of the single (double) Gaussian fit of the detections distribution associated to the interference (which-path) configuration
after removing the background
and renormalizing for the
different detector efficiencies.
The resulting relative frequencies $f_{\pm}$ 
and their Poissonian errors are plotted in Figure~\ref{fig:4} 
for the two satellites.

\begin{figure}[h]
\centering
\includegraphics[width=8cm]{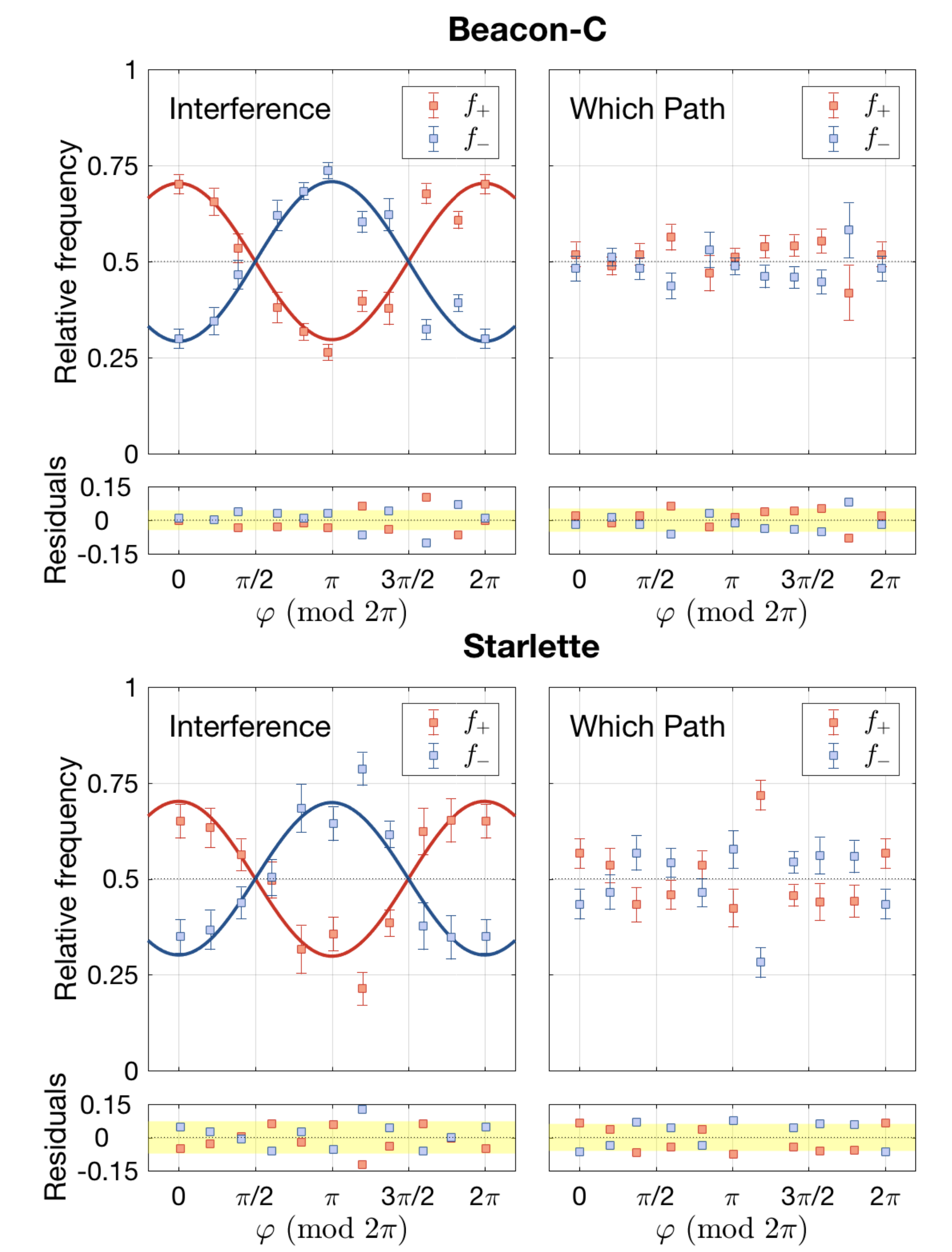}
\caption{ {\it Experimental results for the interference and which-path configurations.} Relative frequencies $f_{\pm}$ of counts in the two detectors $\mathrm{Det}_{\pm}$ as a function of the kinematic phase $\varphi$ introduced by the satellite for the passages of Beacon-C and Starlette satellites. The error bars are estimated using the Poissonian error associated to counts.
Below each plot we show the relative residuals as a function of $\varphi$.  
We note that at the point $\varphi \approx 0$ and $\varphi \approx 2\pi$ the same subset of data was selected. 
In the ``interference'' configuration, we estimated from the fitted data a visibility  $\mathcal{V}^{\mathrm{B}} = 41\pm 4\%$ for Beacon-C and 
$\mathcal{V}^{\mathrm{S}} = 40 \pm 4\%$  for Starlette. } 
\label{fig:4}
\end{figure}

For the ``interference'' subset of the data we 
may observe the relative phase information 
by erasing the photon's ``which-path'' information. This is evident by the recovery of the interference pattern shown in the left part of Figure~\ref{fig:4}.
By fitting the data with $P_{\pm} = (1 \pm \mathcal{V}_{\mathrm{exp}} \cos\varphi)/2$ given by Eq.~\eqref{eq_interference}, we obtained an 
experimental visibility value $\mathcal{V}_{\mathrm{exp}}\approx 40\%$ for both satellites and a clear phase dependent modulation in the two detector outcomes. Furthermore, the visibility obtained during preliminary tests where the sHWP was fixed in the {\bf off} mode, is compatible with the results obtained while performing the delayed choice, attesting that the latter had no influence in the observed interference pattern.
The value of
the experimental visibility, lower than the theoretical value of $100\%$, is due to experimental imperfections in the
MZI and to residual birefringence caused by the telescope Coud\'e. 
This result validates once more the theoretical model for the kinematic phase $\varphi(t)$ introduced and exploited in our recent work~\cite{Vallone_prl_2016}.

On the other hand, the ``which-path'' relative frequencies are constant (within statistical fluctuations) for all va\-lues of $\varphi$, as predicted by the theoretical model $P_{\pm} = 0.5$. 
In this case, the ``which-path'' measurement destroys any information about 
the relative phase of the two time-bins.

When the photon's particle-like nature is inquired, we obtain conclusive which-path information with probability $p_{\rm wp}=91 \pm 1\%$ ($95 \pm 1\%$) for Beacon-C (Starlette). Such values are obtained by 
the ratio between the counts in the lateral peaks and the total ones: indeed, when the photon is detected in one of the two lateral peaks, {\it which-path} information is recovered.
Since classical particles should always give complete which-path information, we could naively conclude that our photons behave as classical particles at least $91\%$ of the time. If such interpretation were correct then we would expect interference with at most $9\%$ visibility when the photon's wave-like nature is inquired. This is in remarkable contrast with the measured visibility, which is 
at least $8\sigma$ distant from that prediction, allowing us to exclude  any model where the photon behaves as a purely classical particle. 

The agreement between the theoretical model and the obtained results can be assessed by calculating the residuals between the fit and the experimental data.  
From Figure~\ref{fig:4}, we can observe that these residuals are randomly distributed within the foreseeable Poissonian fluctuations: indeed most points lay within $\pm 1.5 \sigma$ from the expected values, where $\sigma$ is the mean error. This can also be seen  by calculating the root mean square of the residuals $\sigma_R \sim 0.05$ for both satellites, which is compatible with the expected statistical fluctuations.

Given the optical losses $\eta_{opt}=0.13$ in the receiving setup and the detection efficiency $\eta_{det}=0.1$, the mean number of photons $\mu$ in the received pulses can be derived by measuring the detection rate. At the primary mirror we received $\mu \approx 2.2 \times 10^{-3}$ for Starlette and $\mu \approx 1.9 \times 10^{-3}$ for Beacon-C. From these values we can conclude that  the particle and wave-like properties are measured at the single photon level since the probability of having more than one photon per pulse passing through the MZI on the way back is $\sim \mu^2 /2 ,$ that is at most of the order of $10^{-6}$.

\bigskip
{\it Conclusions.\textemdash}
We realized Wheeler's {\it delayed-choice Gedankenexperiment} along a Space channel involving LEO satellites by combining two independent degrees of freedom of light. 
This result extends the validity of the quantum mechanical description of complementarity to the spatial scale of  LEO orbits. Furthermore it supports the feasibility of  efficient encoding exploiting both polarization and time-bin for high-dimensional quantum key distribution~\cite{ZhongHDQKD}.  
Finally, our work paves the way for satellite implementation of other foundational-like tests 
and applications of quantum mechanics involving hyperentangled states \cite{Cabello_2006,barbieri06prl,Graham2016}, around the planet and beyond. 

\bigskip
{\it Acknowledgments.\textemdash}
We would like to thank Francesco Schiavone, Giuseppe Nicoletti, and the MLRO technical operators for the collaboration and support for their contributions to the setup.
{F.V., L.C., and M.S.
acknowledge the Center of Studies and Activities for Space
(CISAS) ``Giuseppe Colombo'' for financial support.}

Our work was partially supported by the Strategic-Research-Project QUINTET of the Department of Information Engineering, University of Padova, and the Strategic-Research-Project QUANTUMFUTURE of the University of Padova.

\vskip3cm
\onecolumngrid
\appendix
\newpage

\section{Appendix: Experimental details}
The laser pulse train used in the delayed-choice experiment is generated by a Nd:YVO$_4$ mode locking master oscillator paced by an atomic clock and stabilized at the repetition rate of 100 MHz, corresponding to a temporal separation between the pulses of 10 ns.
The 1064 nm pulses are then up-converted to the desired wavelength $\lambda = 532$ nm with a periodically poled lithium niobate crystal. The mean power of the train is of the order of 100 mW, corresponding to an energy per pulse of 1 nJ. The beam is sent through a bulk Mach-Zehnder-Interferometer (MZI)
width an unbalance of about one meter where, in the long arm, two $4f$ relaying optical systems guarantee the matching of the beam wave fronts. To mitigate optical aberration we design each $4f$ system by using two doublets (meniscus and plano convex lens) of equivalent focal length of about 125 cm. 

We measured the temporal unbalance $\Delta t$ of the MZI by sending the pulsed train through it 
ad putting a SPAD (PDM series by Micro Photon Devices) in one of the beam-splitter output port. 
As expected, the detections appear at two different times in two well separated peaks. Each peak is characterized by a exponentially modified Gaussian distribution whose standard deviation is of the order of 40 ps (due to the timing jitter of the detector and the pulse duration). 
By fitting the distribution we estimate the unbalance
of the MZI as $\Delta t = 3.498 \pm 0.002$ ns.

The sHWP is composed by two liquid crystal retarders (LCRs)  mounted with orthogonal axes. Each LCR introduces a phase retardation between the two orthogonal polarization modes of the impinging light, dependent on the voltages applied. We characterized the two LCRs by measuring the birefringence introduced as a function of the applied voltage and then designed the two sHWP to act as a single fast switching half wave plate inclined at $45\degrees$. With this configuration, we obtain a switching time $t_{\mathrm{sHWP}} \lesssim 500 \ \mu$s.

The 100 MHz train is directed to the satellite via a 1.5 m diffraction limited Ritchey–Chrétien telescope telescope designed for laser ranging tracking \cite{mlro,Bianco}. Because of back-reflection
capacity of the corner cubes mounted on the satellites, the MLRO observatory is automatically in the illuminated cone at the ground.

At the MZI-PBS output port the beam is focused and spectrally filtered before passing through a quarter-waveplate (QWP), a half-waveplate (HWP) and the polarizing beam splitter (PBS) which perform the polarization measurement in the $\lbrace \ket{+}, \ket{-} \rbrace$ basis. 
The photons are finally collected by two single photon photomultipliers (detection efficiency $\sim 10\%$, 22 mm diameter) whose detection times are recorded by a time-to-digital converter (QuTau time tagger) with 81 ps resolution. The time tagger also records the encoded value of bit $b$ extracted by the QRNG.  

To determine the expected time of arrival $t_{\mathrm{ref}}$ of the reflected pulses and to estimate the value $\varphi(t)$ of the kinematic phase shift introduced by the satellite, the 100 MHz train is synchronized to a strong 10 Hz laser train used for satellite laser ranging (SLR). The SLR train is generated by the same mode-locking master oscillator used for the 100 MHz train by selecting one pulse every $10^7$. Each SLR pulse is then amplified and up-converted by a SHG stage resulting in a SLR train at 532 nm with 1 W of mean power (corresponding to an energy of 100 mJ per pulse at the repetition rate of 10 Hz). A non polarizing beam splitter is used for combining the two pulsed beams before sending them to the target satellite via the MLRO telescope. 

By taking into account the Doppler effect, we can estimate the instantaneous radial velocity with respect to the ground station $v_r(t) = c(\Delta T' - \Delta T)/(\Delta T' + \Delta T)$ where $\Delta T'$ is the temporal separation of two consecutive SLR pulses in reception, while  $\Delta T= 1/(10 \ \mbox{Hz}) = 100$ ms is the temporal separation of two consecutive SLR pulses in transmission \cite{vallone_spie_2016}.  This information is crucial for estimating the kinematic phase shift introduced by the satellite, since it is continuously changing along the orbit.

\end{document}